\documentclass[journal,dvips]{IEEEtran}

\usepackage[pdf]{pstricks}
\usepackage{pst-plot}
\usepackage{pstricks-add}
\usepackage{overpic}
\usepackage{pst-eps}

\usepackage[latin1]{inputenc}
\usepackage{graphicx} 
\usepackage{amsmath} 
\usepackage{amssymb}  
\usepackage{mathtools}
\mathtoolsset{showonlyrefs}
\usepackage{flushend}
\usepackage{url}
\usepackage{color}

\IEEEoverridecommandlockouts

\def\vec#1{\underline{#1}} 
\def\mat#1{{\mathbf #1}} 
\def\E{\mathrm{E}} 

\mathtoolsset{showonlyrefs}

\begin{document}

\author{Sebastian~Dingler$^1$ (orcid: 0000-0002-0162-8428) \thanks{$^1$This paper is an independent contribution without affiliation. To mail me remove foobar \tt\small s.dinglerfoobar@gmail.com}  }

\title{State estimation with the Interacting Multiple Model (IMM) method}

\maketitle

\begin{abstract}
For model-based estimation methods, the modeling that is as close to reality as possible makes a vital estimation result.
In simple applications, it is sufficient to model a system with a single state-space model.
However, there are applications in which a system changes its behavior deterministically or stochastically.
A previously defined model then describes the behavior of the system only inaccurately or is even no longer valid.
The state of the art is to use more than one system model in parallel and to derive a suitable system estimate from them.
In the literature, this is generally referred to as the Multiple Model (MM) method.
Depending on the application and requirements, different methods exist for this purpose, which determines a single state estimate from a set of models.
A frequently used representative of these methods is the Interacting Multiple Model (IMM) method which will be presented in this paper.
\end{abstract}


\section{Introduction}
\IEEEPARstart{I}{n} the framework of Bayesian estimation theory, the Kalman filter is the most important representative for estimating the state of a linear stochastic system \cite{Kalman.1960}. 
If the minimum variance is used as the optimality criterion, no other linear method gives a better result than the Kalman filter. 
The description of linear stochastic systems is done with the state-space representation. Setting up a suitable state-space model is not always a trivial process. 
An additional challenge occurs when systems have different modes of operation. 
In addition to the actual model building, the task of determining when which model is valid must be solved. 
A concrete example is the observation of an automobile. 
A car obeys different physical laws when it drives through a curve than when it merely drives straight ahead. For the external observer, however, it may be unknown when these different maneuvers are performed. 
A single model would not be sufficient for this estimation task. 
Even more, if a model does not fit the behavior of a system, the estimation result may even diverge.

The solution is to represent all the different operating modes of a system with several state-space models.
Subsequently, it must be determined in which mode the system is currently in.
For this purpose, different methods have been developed to select, combine and link the different models.
In the literature these procedures are called the Multiple Model (MM) method \cite{Li.2005}.
In this paper, a specific method, the Interacting Multiple Model (IMM) method, will be discussed in detail. 
The IMM method is widely used because it represents a good compromise between complexity and performance \cite{BarShalom.2001,Blackman.1999}.

Before going into more detail about the IMM method, Section \ref{sec:fundamentals} takes a look at the fundamentals and the historical development of the MM method.
Subsequently, the IMM method is discussed in Section \ref{sec:imm}.
The operation is illustrated by a simulation in Section \ref{sec:simulation}.
Finally, an outlook on the extensions of the MM methods is given in Section \ref{sec:summary}.

\section{Fundamentals}
\label{sec:fundamentals}
Linear stochastic systems are usually described using the state-space representation. 
This is in continuous-time a chain of first-order differential equations. 
Since in practice state estimation is discrete-time, systems are described with the following difference equation
\begin{equation}
\vec{x}_{k+1}=\mat{A}_k  \vec{x}_{k} + \mat{B}_k \vec{u}_k + \vec{w}_k \enspace .
\end{equation}
Here the internal quantities, the state variables, are combined to the state vector $\vec{x}$.
The system matrix $\mat{A}_k$ maps the system state at time $k$ to the system state at time $k+1$.
Optionally, there is an addition of the product of the input matrix $\mat{B}_k$ and the control variable $\vec{u}_k$ when external variables act on the system.
In addition, there is an additive vectorial white noise $\vec{w}_k$.
This is assumed to be Gaussian distributed with $\vec{w}_k \sim \mathcal{N}(0,\mat{Q}_{k})$.
In general, the state $\vec{x}_{k}$ is not directly accessible, but can only be determined via the following mapping
\begin{equation}
\vec{y}_{k}=\mat{H}_k  \vec{x}_{k} + \vec{e}_k \enspace .
\end{equation}
The measurement matrix $\mat{H}_k$ projects the state-space onto the space of measured values $\vec{y}_{k}$. 
The measured values additionally experience additive vectorial white noise $\vec{e}_k$, which is assumed to be Gaussian distributed $\vec{e}_k \sim \mathcal{N}(0,\mat{R}_{k})$. 
The filter presented by Kalman provides an estimate $\vec{\hat{x}}_{k}$ of the state $\vec{x}_{k}$ based on the measurements $\vec{y}_{1:k}=\left\{\vec{y}_1,\vec{y}_2,...,\vec{y}_k\right\}$ available up to time $k$.
Since $\vec{w}_k$ and $\vec{e}_k$ are Gaussian distributed random variables and the system is linear, the first two moments
\begin{equation}
\hat{\vec{x}}_{k|k}=\E\left\{\vec{x}_k|\vec{y}_{1:k}\right\}
\end{equation}
and
\begin{equation}
\mat{C}_{k|k}=\E\left\{ \left(\vec{x}_k -\hat{\vec{x}}_{k|k} \right)\left(\vec{x}_k -\hat{\vec{x}}_{k|k} \right)^T |\vec{y}_{1:k}\right\}
\end{equation}
suffice to uniquely describe the state of the system. 
The Kalman filter equations are divided in the prediction step  (marked with the subscript $k+1|k$)
\begin{align}
\vec{\hat{x}}_{k+1|k} &=\mat{A}_{k} \vec{\hat{x}}_{k|k} + \mat{B}_{k} \vec{u}_{k} \\
\mat{C}_{k+1|k} &=\mat{A}_{k} \mat{C}_{k|k} (\mat{A}_{k})^T + \mat{B}_{k} \mat{Q}_{k} (\mat{B}_{k})^T
\end{align}
and into the filter step (marked with the subscript $k|k$)
\begin{align}
\mat{K}_{k} &=\mat{C}_{k|k-1} (\mat{H}_{k})^T (\mat{R}_{k} + \mat{H}_{k} \mat{C}_{k|k-1} (\mat{H}_{k})^T)^{-1} \\
\vec{\hat{x}}_{k|k} &=\vec{\hat{x}}_{k|k-1}+\mat{K}_{k}(\vec{y}_k-\mat{H}_{k}\vec{\hat{x}}_{k|k-1}) \\
\mat{C}_{k|k} &=\mat{C}_{k|k-1}-\mat{K}_{k}\mat{H}_{k}\mat{C}_{k|k-1} \enspace .
\end{align}
Important for the following explanations are the auxiliary variables innovation
\begin{align}
\vec{\tilde{y}}_k & =\vec{y}_k-\mat{H}\vec{\hat{x}}_{k|k-1} \\
									& =\vec{y}_k-\vec{\hat{y}}_{k|k-1}
\end{align}
and residual covariance
\begin{equation}
\mat{S}_k=\mat{R}_{k} + \mat{H}_{k} \mat{C}_{k|k-1} (\mat{H}_{k})^T \enspace .
\end{equation}
The innovation describes the error made by the prediction step of the Kalman filter.
The residual covariance is the covariance of the prediction error.
These auxiliary variables are important because they can be used to determine how well the system model fits the real system behavior.
\section{The Multiple Model (MM) Method}
The MM methods, make use of $r$ models $m^{(i)} \in \mathbb{M}=\left\{m^{(1)},m^{(2)},...,m^{(r)}\right\}$ to represent a system. 
Therefore the system description results to 
\begin{equation}
\vec{x}_{k+1}=\mat{A}^{(i)}_k  \vec{x}_{k} + \mat{B}^{(i)}_k \vec{u}_k + \vec{w}^{(i)}_k
\end{equation}
with
\begin{equation}
\mat{A}^{(i)}_k = \mat{A}(m^{(i)}_k)\enspace, \mat{B}^{(i)}_k = \mat{B}(m^{(i)}_k)\enspace, \vec{w}^{(i)}_k = \vec{w}(m^{(i)}_k)\enspace .
\end{equation}
It should be emphasized that the set $\mathbb{M}$ is time-invariant. This means that the number of possible models does not change over time. Two questions now arise:
\begin{enumerate}
	\item According to which model does the system behave?
	\item If so and when is there a switch between models?
\end{enumerate}
Various methods have been developed to answer these questions that make different assumptions and simplifications.
The first work on this was done only a few years after Kalman presented his filter.
Numerous works followed until today, allowing more complex behaviors when changing the models.
In a detailed work of Li which gives an overview of the MM methods, the MM methods are divided into three generations \cite{Li.2005}. 
Thereby the first generation deals only with the first question. 
The second-generation (to which the IMM method belongs) also provides an answer to the second question. 
Since the generations build on each other, the first generation will be discussed in the following.

\subsection{First generation: Autonomous Multiple Model (AMM) method}
In the first generation MM method, the constraint is imposed that the system is in a single mode at each time $k$. The optimal state estimation is achieved exclusively with only one model
\begin{equation}
m^{(i)}=m^{(i)}_k, m^{(i)} \in \mathbb{M} \enspace .
\end{equation} 
We are looking for the model $m^{(i)}$ that is not known a priori. The solution is to use $r$ parallel elementary Kalman filters that provide a state estimate $\left\{\vec{\hat{x}}^{(i)}_{k|k},\mat{C}^{(i)}_{k|k}\right\}$ for each model $m^{(i)}$. From these elementary estimates, the law of total expected values is used to find the optimal estimate $\vec{\hat{x}}_{k|k}$ for $\vec{x}_k$
\begin{equation}
\vec{\hat{x}}_{k|k}=\sum^{r}_{i=1} \mu^{(i)}_k \vec{\hat{x}}^{(i)}_{k|k} \enspace,
\end{equation}
where $\mu^{(i)}_k = P(m^{(i)}|\vec{y}_{1:k})$ is the posteriori probability that an optimal single estimate is obtained with model $m^{(i)}$. 
Unbiasedness is achived by $\sum^{r}_{i=1} \mu_i=1$. 
The covariance of $\vec{\hat{x}}_{k|k}$ is given by
\begin{equation}
\mat{C}_{k|k}=\sum^{r}_{i=1} \mu^{(i)}_{k} \left\{\mat{C}^{(i)}_{k|k}+(\vec{\hat{x}}^{(i)}_{k|k}-\vec{\hat{x}}_{k|k})(\vec{\hat{x}}^{(i)}_{k|k}-\vec{\hat{x}}_{k|k})^T\right\}  \enspace .
\end{equation}
The probability $P(m^{(i)}|\vec{y}_{1:k})$ that $m^{(i)}$ is the model associated with the system can be determined using Bayes' theorem
\begin{align}
  P(m^{(i)}|\vec{y}_{1:k}) &= P(m^{(i)}|\vec{y}_k,\vec{y}_{1:k-1}) \\
        &= \frac{P(m^{(i)}|\vec{y}_{1:k-1})f(\vec{y}_k|\vec{y}_{1:k-1},m^{(i)})}{P(\vec{y}_k|\vec{y}_{1:k-1})} \\
        &= \frac{P(m^{(i)}|\vec{y}_{1:k-1})f(\vec{y}_k|\vec{y}_{1:k-1},m^{(i)})}{\sum^{r}_{j=1} P(m^{(j)}|\vec{y}_{1:k-1})f(\vec{y}_k|\vec{y}_{1:k-1},m^{(j)})} \enspace .
\end{align}
With $\mu^{(i)}_{k-1}=P(m^{(i)}|\vec{y}_{1:k-1})$ and $L^{(i)}_k=f(\vec{y}_k|\vec{y}_{1:k-1},m^{(i)})$ we get the recursion
\begin{equation} \label{eq:rec}
  \mu^{(i)}_k =\frac{\mu^{(i)}_{k-1} L^{(i)}_k}{\sum^{r}_{j=1} \mu^{(j)}_{k-1} L^{(j)}_k} \enspace .
\end{equation}
Here $L^{(i)}_k$ is the probability distribution that describes whether to use the model $m^{(i)}$. This is Gaussian distributed 
\begin{equation}
  L^{(i)}_k =\mathcal{N}(\vec{y}_k;\vec{\hat{y}}^{(i)}_{k|k-1},\mat{S}^{(i)}_{k})=\mathcal{N}(\vec{y}_k-\vec{\hat{y}}^{(i)}_{k|k-1};0,\mat{S}^{(i)}_{k}) \enspace .
\end{equation}
An important property of the recursion equation \eqref{eq:rec} is that it converges to $1$ for the correct model $m^{(i)}$. This has the consequence that the system is estimated exclusively with the model that best describes the system behavior.

\section{The Interacting Multiple Model (IMM) Method}
\label{sec:imm}
The first generation of MM methods only allows the possibility to describe a system for all time points $k$ with one single model. As already stated, in practice systems are of interest which change their behavior randomly from the observer's point of view. The second-generation of MM methods allows such behavior. Here, the change between models is modeled as a Markov process. However, this freedom leads to the problem that for each possible sequence of models a Kalman filter is needed to obtain an optimal result. For $n$ time steps, this means $r^n$ filters. Each estimation result would have to be weighted fused from $r^n$ possible combinations. For practical purposes, this is too many possible combinations. To address this problem, suboptimal procedures have been developed that attempt to keep the number of Kalman filters at $r$ or as close to $r$ as possible. One of these methods is the IMM method, which gets by with $r$ filters.

The derivation is similar to the first generation, but differs in some points. Thus the estimation result for the state, from the elementary Kalman filters, results analogously to the first generation with
\begin{equation}
\vec{\hat{x}}_{k|k}=\sum^{r}_{i=1} \mu^{(i)}_{k} \vec{\hat{x}}^{(i)}_{k|k}
\end{equation}
and
\begin{equation}
\mat{C}_{k|k}=\sum^{r}_{i=1} \mu^{(i)}_{k} \left\{\mat{C}^{(i)}_{k|k}+(\vec{\hat{x}}_{k|k}-\vec{\hat{x}}^{(i)}_{k|k})(\vec{\hat{x}}_{k|k}-\vec{\hat{x}}^{(i)}_{k|k})^T\right\} \enspace .
\end{equation}
Here $\mu^{(i)}_k = P(m^{(i)}_k|\vec{y}_{1:k})$ is the posteriori probability that the model $m^{(i)}_k$ yields the optimal single estimate at time $k$. Analogous to the first generation, the probability $P(m^{(i)}_k|\vec{y}_{1:k})$ can be calculated using Bayes' theorem
\begin{align}
  P(m^{(i)}_k|\vec{y}_{1:k}) &= P(m^{(i)}_k|\vec{y}_k,\vec{y}_{1:k-1}) \\
        &= \frac{P(m^{(i)}_k|\vec{y}_{1:k-1})f(\vec{y}_k|\vec{y}_{1:k-1},m^{(i)}_k)}{P(\vec{y}_k|\vec{y}_{1:k-1})} \\
        &= \frac{P(m^{(i)}_k|\vec{y}_{1:k-1})f(\vec{y}_k|\vec{y}_{1:k-1},m^{(i)}_k)}{\sum^{r}_{j=1} P(m^{(j)}_k|\vec{y}_{1:k-1})f(\vec{y}_k|\vec{y}_{1:k-1},m^{(j)}_k)} \enspace .
\end{align}
Here applies again
\begin{align}
  L^{(i)}_k & = f(\vec{y}_k|\vec{y}_{1:k-1},m^{(i)}_k) \\
  & = \mathcal{N}(\vec{y}_k;\vec{\hat{y}}^{(i)}_{k|k-1},\mat{S}^{(i)}_{k}) \\
  & =\mathcal{N}(\vec{y}_k-\vec{\hat{y}}^{(i)}_{k|k-1};0,\mat{S}^{(i)}_{k}) \enspace .
\end{align}
With the addition of the law of total probability
\begin{align}
P(m^{(i)}_k|\vec{y}_{1:k-1}) &= \sum^r_{j=1} P(m^{(i)}_k|m^{(j)}_{k-1},\vec{y}_{1:k-1}) P(m^{(j)}_{k-1}|\vec{y}_{1:k-1}) \enspace .
\end{align}
Here $P(m^{(i)}_k|m^{(j)}_{k-1},\vec{y}_{1:k-1})$ is the transition probability from model $m^{(j)}_{k-1}$ at time $k-1$ to model $m^{(i)}_k$ at time $k$. This corresponds to the transition probability for Markov chains and is abbreviated by 
\begin{equation}
p_{ji}=P(m^{(i)}_k|m^{(j)}_{k-1},\vec{y}_{1:k-1}) \enspace .
\end{equation}
Since $\mu^{(i)}_{k-1} = P(m^{(i)}_{k-1}|\vec{y}_{1:k-1})$ holds, this can be interpreted as a prediction of the model likelihood
\begin{equation}
\mu^{(i)}_{k|k-1}=P(m^{(i)}_k|\vec{y}_{1:k-1})=\sum^r_{j=1} p_{ji} \mu^{(j)}_{k-1} \enspace .
\end{equation}
In other words, knowing the transition probability from one model to another, it is possible to calculate how likely it is that a given model will be valid at the next time point. 
The complete recursion equation for this is
\begin{equation}
\mu^{(i)}_k=\frac{\mu^{(i)}_{k|k-1}L^{(i)}_k}{\sum^r_{j=1} \mu^{(j)}_{k|k-1}L^{(j)}_k} \enspace .
\end{equation}
Since the IMM method always considers the case of a possible model change, the knowledge about the model likelihood is used to calculate the $r$ Kalman filters with
\begin{equation}
\vec{\bar{x}}^{(i)}_{k-1|k-1}=\sum^{r}_{j=1} \vec{\hat{x}}^{(j)}_{k-1|k-1} \mu^{(j|i)}_{k-1}
\end{equation}
and
\begin{multline}
\bar{\mat{C}}^{(i)}_{k-1|k-1}=\sum^{r}_{j=1} \mu^{(j|i)}_{k-1} \left\{\mat{C}^{(j)}_{k-1|k-1} \right.\\
\left.+(\vec{\bar{x}}^{(i)}_{k-1|k-1}-\vec{\hat{x}}^{(j)}_{k-1|k-1})(\vec{\bar{x}}^{(i)}_{k-1|k-1}-\vec{\hat{x}}^{(j)}_{k-1|k-1})^T\right\}
\end{multline}
to be initialized individually. The weight $\mu^{(j|i)}_{k-1}$ describes the probability that the model $m^{(i)}$ was correct at time $k-1$, given the model $m^{(j)}$ will be valid at time $k$. This is done by an inference using Bayes' theorem
\begin{align}
\mu^{(j|i)}_{k-1} &= P(m^{(j)}_{k-1}|m^{(i)}_k,\vec{y}_{1:k-1}) \\
                  &= \frac{P(m^{(i)}_k|m^{(j)}_{k-1},\vec{y}_{1:k-1}) P(m^{(j)}_{k-1}|\vec{y}_{1:k-1})}{P(m^{(i)}_k|\vec{y}_{1:k-1})} \\
                  &= \frac{p_{ji} \mu^{(j)}_{k-1}}{\mu^{(i)}_{k|k-1}} \enspace .
\end{align}

This initialization can then be used to determine the new individual estimates using the Kalman filter equations. Table \ref{tab:imm-cycle} shows a complete cycle of the IMM method in summary.

\begin{table}[ht]
 \caption{One cycle of the IMM method}
 \label{tab:imm-cycle}

 		\begin{tabular}{ l  }
 		 \hline   
       Initialization of $r$ elementary Kalman filters:  \\
$\mu^{(i)}_{k|k-1}=\sum^r_{j=1} p_{ji} \mu^{(j)}_{k-1}$ \\
 $\mu^{(j|i)}_{k-1} =  \frac{p_{ji} \mu^{(j)}_{k-1}}{\mu^{(i)}_{k|k-1}}$ \\
 
 			 $\vec{\bar{x}}^{(i)}_{k-1|k-1}=\sum^{r}_{j=1} \vec{\hat{x}}^{(j)}_{k-1|k-1} \mu^{(j|i)}_{k-1}$ \\

$\begin{aligned}
\bar{\mat{C}}^{(i)}_{k-1|k-1} &=\sum^{r}_{j=1} \mu^{(j|i)}_{k-1} \left\{\mat{C}^{(j)}_{k-1|k-1} \right.\\
& \left.+(\vec{\bar{x}}^{(i)}_{k-1|k-1}-\vec{\hat{x}}^{(j)}_{k-1|k-1})(\vec{\bar{x}}^{(i)}_{k-1|k-1}-\vec{\hat{x}}^{(j)}_{k-1|k-1})^T\right\}
\end{aligned}$ \\

		State estimation with $r$ parallel Kalman filters:\\
			 $\vec{\hat{x}}^{(i)}_{k+1|k} =\mat{A}^{(i)}_{k} \vec{\bar{x}}^{(i)}_{k|k} + \mat{B}^{(i)}_{k} \vec{u}^{(i)}_{k}$\\
			 			 $\mat{C}^{(i)}_{k+1|k} =\mat{A}^{(i)}_{k} \bar{\mat{C}}^{(i)}_{k|k} (\mat{A}^{(i)}_{k})^T + \mat{B}^{(i)}_{k} \mat{Q}^{(i)}_{k} (\mat{B}^{(i)}_{k})^T$\\
			 
			 $\mat{S}^{(i)}_{k} =\mat{R}^{(i)}_{k} + \mat{H}^{(i)}_{k} \mat{C}^{(i)}_{k|k-1} (\mat{H}^{(i)}_{k})^T$ \\
			 
			$\mat{K}^{(i)}_{k} =\mat{C}^{(i)}_{k|k-1} (\mat{H}^{(i)}_{k})^T (\mat{S}^{(i)}_{k})^{-1}$ \\
			
			$\vec{\hat{x}}^{(i)}_{k|k} =\vec{\hat{x}}^{(i)}_{k|k-1}+\mat{K}^{(i)}_{k}(\vec{y}_k-\mat{H}^{(i)}_{k}\vec{\hat{x}}^{(i)}_{k|k-1})$ \\
			 
		$\mat{C}^{(i)}_{k|k} =\mat{C}^{(i)}_{k|k-1}-\mat{K}^{(i)}_{k}\mat{H}^{(i)}_{k}\mat{C}^{(i)}_{k|k-1}$ \\

		Updating the model likelihood:  \\
		  $L^{(i)}_k =\mathcal{N}(\vec{y}_k-\vec{\hat{y}}^{(i)}_{k|k-1};0,\mat{S}^{(i)}_{k})$ \\
		$\mu^{(i)}_k=\frac{\mu^{(i)}_{k|k-1}L^{(i)}_k}{\sum^r_{j=1} \mu^{(j)}_{k|k-1}L^{(j)}_k}$ \\
		
		Total state estimation from $r$ elementary Kalman filters: \\

$\vec{\hat{x}}_{k|k}=\sum^{r}_{i=1} \mu^{(i)}_{k} \vec{\hat{x}}^{(i)}_{k|k}$ \\

$\mat{C}_{k|k}=\sum^{r}_{i=1} \mu^{(i)}_{k} \left\{\mat{C}^{(i)}_{k|k}+(\vec{\hat{x}}_{k|k}-\vec{\hat{x}}^{(i)}_{k|k})(\vec{\hat{x}}_{k|k}-\vec{\hat{x}}^{(i)}_{k|k})^T\right\}$ \\

 \hline   
		\end{tabular}

 \end{table}

\section{Simulation of the IMM method}
\label{sec:simulation}
To illustrate how the IMM method works, a simulation follows next. 
For this purpose, the following two models were chosen as simulation scenario
\begin{enumerate}
	\item Constant Velocity Modell (CV) with system noise $(\sigma^{(CV)}_w)^2=1$
	\item Constant Acceleration Modell  (CA) with system noise $(\sigma^{(CA)}_w)^2=1$.
\end{enumerate}
The corresponding state-space models can be found in the appendix~\ref{sec:appendix}. 
As a measurement mapping the simulation uses
\begin{align}
\mat{H}^{(CV)}_k = \left[
\begin{array}[l]{cc}
1 & 0\\
\end{array}
\right],\quad
\mat{H}^{(CA)}_k = \left[
\begin{array}[l]{ccc}
1 & 0 & 0\\
\end{array}
\right] .
\end{align}
The transition matrix states
\begin{equation}
p_{ji} = \left[
\begin{array}[l]{cc}
0.75 & 0.25 \\
0.25 & 0.75
\end{array}
\right] \enspace .
\end{equation}
In Fig. \ref{fig:modeProbability}, the possible change between models is plotted (gray dotted line). 
To evaluate the quality of the state estimation with the IMM method, the widely used quality measures Root Mean Squared Error (RMSE) and Normalized Estimation Error Squared (NEES) were chosen \cite{BarShalom.2001}. 
To ensure statistical significance, the simulation was performed with 1000 runs.
\subsection{Results of simulation}
\begin{figure}[ht]
	\psset{xunit=0.035cm,yunit=3.4cm,mathLabel=false}
	\begin{pspicture}(-40,-0.25)(240,1.1)
		\readdata{\modeProbability}{modeProbability.csv}
		\readdata{\mode}{mode.csv}
		\listplot[plotstyle=line,linecolor=black,linewidth=0.8pt,linestyle=solid,yStart=0,yEnd=12]%
						{\modeProbability}
		\listplot[plotstyle=line,linecolor=darkgray,linewidth=1.5pt,linestyle=dotted,dotsep=0.1pt,yStart=0,yEnd=12]%
						{\mode}

		\psaxes[labelFontSize=\footnotesize,Ox=0,Dx=20,Dy=0.1,Oy=0.0](0,0)(0,0)(200,1.05)
		\psline[]{->}(0,1)(0,1.1)
		\psline[]{->}(200,0)(210,0)
		
		\rput(100,-0.19){\footnotesize {Zeitschritte $k$}}
		\rput{90}(-25,0.5){\footnotesize {Wahrscheinlichkeit}}
		\rput(-28,0){\footnotesize {CV=}}
		\rput(-28,1){\footnotesize {CA=}}
		\rput(160,0.8){
			\psframe[linewidth=0.5pt,linecolor=black,fillstyle=solid](0,0)(60,0.16)

			\psline[linecolor=darkgray,linewidth=2pt,linestyle=dotted,dotsep=0.4pt](5,0.12)(15,0.12)
			\rput[l](17,0.12){\sffamily \tiny Wahrer Modus}
			\psline[linecolor=black,linewidth=2pt](5,0.05)(15,0.05)
			\rput[l](17,0.05){\sffamily \tiny $\mu_k^{(CA)}$}
		}
	
		\end{pspicture}
\caption{Probability for the models CV \& CA after 1000 runs.}
\label{fig:modeProbability}
\end{figure}
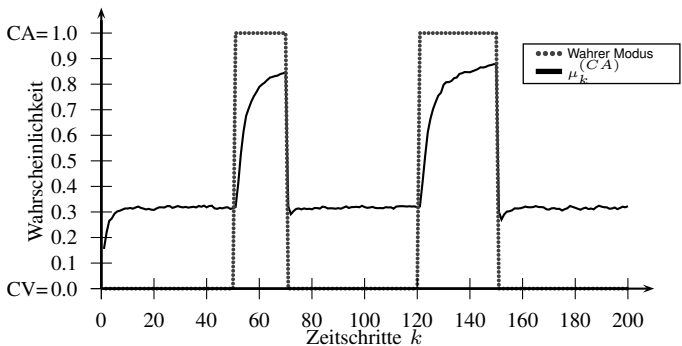

In Fig. \ref{fig:modeProbability}, the estimated model probability $\mu^{(CA)}_k$ for the CA system model is plotted (black line). 
It can be seen that as soon as a transition from the CV to the CA model occurs, the probability for the CA model increases steadily. 
As soon as the system model is transitioned from the CA model to the CV model, it decreases to the original probability level. 
The model probability never becomes zero or one (in contrast to the first MM generation), since a model change is always possible (due to the modeled transition matrix).

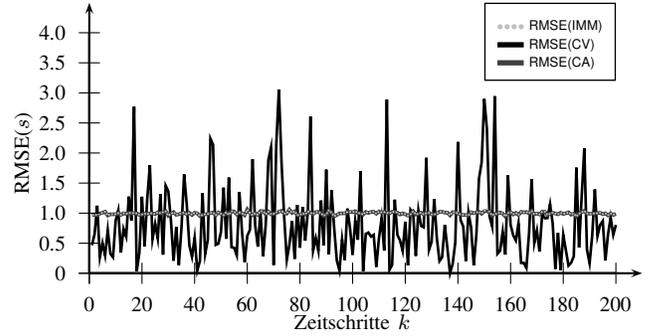
\begin{figure}[ht]
	\psset{xunit=0.035cm,yunit=0.8cm,mathLabel=false}
	\begin{pspicture}(-40,-1)(240,5)
		\readdata{\RMSEIMM}{RMSEIMM.csv}
		\readdata{\RMSEFilterONE}{RMSEFilter1.csv}
		\readdata{\RMSEFilterTWO}{RMSEFilter2.csv}
	
		\listplot[plotstyle=line,linecolor=black,linewidth=1.0pt,linestyle=solid,yStart=0,yEnd=4]%
						{\RMSEFilterONE}
			\listplot[plotstyle=line,linecolor=darkgray,linewidth=1.4pt,linestyle=solid,yStart=0,yEnd=20]%
						{\RMSEIMM}
		\listplot[plotstyle=line,linecolor=lightgray,linewidth=0.8pt,linestyle=dotted,dotsep=0.4pt,yStart=0,yEnd=20]%
						{\RMSEFilterTWO}

		\psaxes[labelFontSize=\footnotesize,Ox=0,Dx=20,Dy=0.5,Oy=0](0,0)(0,0)(200,4)
		\psline[]{->}(0,4)(0,4.5)
		\psline[]{->}(200,0)(210,0)
		
		\rput(100,-0.8){\footnotesize {Zeitschritte $k$}}
		\rput{90}(-25,2){\footnotesize {RMSE($s$)}}
		
		\rput(150,3.2){
			\psframe[linewidth=0.5pt,linecolor=black,fillstyle=solid](0,0)(50,1.3)

			\psline[linecolor=lightgray,linewidth=2pt,linestyle=dotted,dotsep=0.4pt](5,0.9)(15,0.9)
			\rput[l](17,0.9){\sffamily \tiny RMSE(IMM)}
			\psline[linecolor=black,linewidth=2pt](5,0.6)(15,0.6)
			\rput[l](17,0.6){\sffamily \tiny RMSE(CV)}
			\psline[linecolor=darkgray,linewidth=2pt](5,0.3)(15,0.3)
			\rput[l](17,0.3){\sffamily \tiny RMSE(CA)}
		}
	
		\end{pspicture}
\caption{RMSE of the position $s$ after 1000 runs.}
\label{fig:rmse}
\end{figure}

\begin{figure}[ht]
	\psset{xunit=0.035cm,yunit=0.30cm,mathLabel=false}
	\begin{pspicture}(-30,-2.5)(250,12)
		\readdata{\NEESIMM}{NEESIMM.csv}
		\readdata{\NEESFilterONE}{NEESFilter1.csv}
		\readdata{\NEESFilterTWO}{NEESFilter2.csv}
	
		
		\listplot[plotstyle=line,linecolor=lightgray,linewidth=0.8pt,linestyle=solid,yStart=0,yEnd=12]%
						{\NEESFilterONE}
	\listplot[plotstyle=line,linecolor=darkgray,linewidth=0.8pt,linestyle=solid,yStart=0,yEnd=12]%
					{\NEESFilterTWO}
		
	\listplot[plotstyle=line,linecolor=black,linewidth=0.8pt,linestyle=dashed,yStart=0,yEnd=10]
					{\NEESIMM}
\psline[linecolor=lightgray,linewidth=0.5pt,linestyle=dashed](0,2.1258)(200,2.1258)
\psline[linecolor=lightgray,linewidth=0.5pt,linestyle=dashed](0,1.8779)(200,1.8779)

\psline[linecolor=darkgray,linewidth=0.5pt,linestyle=dashed](0,3.1537)(200,3.1537)
\psline[linecolor=darkgray,linewidth=0.5pt,linestyle=dashed](0,2.8501)(200,2.8501)

		\psaxes[labelFontSize=\footnotesize,Ox=0,Dx=20,Dy=1,Oy=0](0,0)(0,0)(200,11.5)
		\psline[]{->}(0,11)(0,12)
		\psline[]{->}(200,0)(210,0)
		
		\rput(100,-2.2){\footnotesize {Zeitschritte $k$}}
			\rput{90}(-20,5){\footnotesize {NEES}}
		
		\rput(155,10){
			\psframe[linewidth=0.5pt,linecolor=black,fillstyle=solid](0,0)(50,1.7)

			\psline[linecolor=black,linewidth=2pt,linestyle=dashed](5,1.3)(15,1.3)
			\rput[l](17,1.3){\sffamily \tiny NEES(IMM)}
			\psline[linecolor=lightgray,linewidth=2pt](5,0.8)(15,0.8)
			\rput[l](17,0.8){\sffamily \tiny NEES(CV)}
			\psline[linecolor=darkgray,linewidth=2pt](5,0.3)(15,0.3)
			\rput[l](17,0.3){\sffamily \tiny NEES(CA)}
		}
	
		\end{pspicture}
\caption{NEES after 1000 runs with 95\% confidence interval [1.87; 2.12] and [2.85; 3.15], respectively.}
\label{fig:nees}
\end{figure}
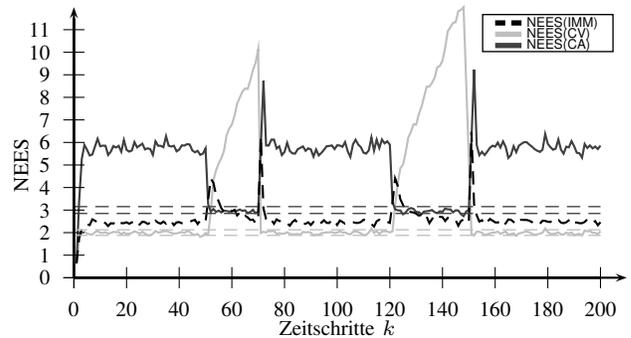

In Fig. \ref{fig:rmse} the RMSE of the position $s$ of the IMM estimate can be seen (gray dotted line). Additionally, the corresponding elementary single estimates of the Kalman filters for the CV and CA model are shown. It is obvious that the IMM estimate is minimally lower than the RMSE(CA) value. The reason is that the CV model is a special case of the CA model. However, this does not mean that the CA model would be sufficient for this simulation scenario. This is shown by the following NEES test (see Fig.~\ref{fig:nees}).

If the NEES of the elementary Kalman filters is considered, it can be seen that they only estimate the error of their own estimation correctly if the corresponding mode is active. Otherwise, the elementary filters estimate the error lower than it actually is (NEES is above the confidence interval). The NEES for the IMM estimate, on the other hand, is below the confidence interval for most time points. This means that the IMM estimate estimates its own estimation error to be larger than it actually is. This corresponds to a conservative estimation behavior that is less problematic for practice than the reverse case. Since NEES(IMM) is not within the confidence interval, it is also evident that the IMM procedure is a suboptimal estimation procedure.

\section{Summary and outlook}
\label{sec:summary}
If systems run through different operating modes and the state is to be estimated as accurately as possible, it makes sense to use a separate state-space model for each mode. 
The task of MM methods is to estimate the current operating mode and to deliver the best possible estimate for the state. 

For the first generation, it was characteristic that it was assumed that the system behaves according to only one model out of a set. 
For this purpose, one elementary Kalman filter was used for each model. 
Then, based on the innovation, it was chosen which filter estimates the system best and its estimation result was used. 
This estimation result is optimal in the mathematical sense with the assumptions mentioned above.

The second-generation of MM methods (which includes the IMM method) assumes that the system can randomly switch operating modes. 
This model switching is described by a homogeneous Markov chain. 
To obtain an optimal estimation result here $r^n$ elementary filters would be necessary, which is not feasible in practice. 
Therefore, there are various suboptimal
methods that use fewer elementary filters. An important representative of these methods is the IMM method, which was presented here. 
The IMM strategy gets along with $r$ filters by merging the individual estimates and reinitializing each filter intelligently.

A closer look at the equations reveals that this is similar to sensor data fusion. 
However, MM methods are different from sensor data fusion in two aspects. 
First, the system always behaves according to a single model. In this case, only one estimation result is correct. 
In sensor data fusion, all estimates are correct. 
Second, in the MM method, all elementary filters use the same information source. 
However, in sensor data fusion, the information is generated by different sensors.
	
In this work, only linear stochastic systems were considered. 
However, for nonlinear stochastic ones, extensions with Extended Kalman Filter (EKF) and Unscented Kalman Filter (UKF) exist. 
This leads to the IMM-EKF and IMM-UKF methods, respectively.

\appendix[Models used for the simulation]
\label{sec:appendix}
For a detailed discussion of the CV or CA model, the reader is referred to \cite{BarShalom.2001}. Both models have in common that only a white noise $\vec{w}_k$ acts on the system. The state-space model reduces to
\begin{align}
\vec{x}_{k+1}=\mat{A}_k \vec{x}_k + \mat{B}_k \vec{w}_k \enspace.
\end{align}
The only difference is the choice of state variables. The constant velocity model uses the velocity $v$ and the position $s$ as state variables. The state vector is thus
\begin{align}
\vec{x}^{(CV)}_k=
\left[
\begin{array}[l]{c}
s_k\\
v_k
\end{array}
\right]\enspace.
\end{align}
It is assumed that the velocity $v$ is constant from time $k$ to time $k+1$. This leads to the system and input matrix
\begin{align}
\mat{A}^{(CV)}_k=\left[
\begin{array}[l]{cc}
1 & T\\
0 & 1
\end{array}
\right],\quad \mat{B}^{(CV)}_k=\left[
\begin{array}[l]{c}
\frac{1}{2} T^2\\
T
\end{array}
\right]\enspace.
\end{align}
The covariance matrix of the process noise is given by
\begin{align}
\mat{Q}^{(CV)}_k &=\E\left\{\mat{B}^{(CV)}_k\vec{w}_k\vec{w}_k(\mat{B}^{(CV)}_k)^T\right\} \\
          &= \mat{B}^{(CV)}_k (\sigma^{(CV)}_w)^2 (\mat{B}^{(CV)}_k)^T\\
          &=
          \left[
\begin{array}[l]{cc}
\frac{1}{4} T^4 & \frac{1}{3} T^3\\
\frac{1}{3} T^3 & T^2
\end{array}
\right]
(\sigma^{(CV)}_w)^2 \enspace.
\end{align}
The state of the constant acceleration model, is extended by the acceleration $a$
\begin{align}
\vec{x}^{(CA)}_k=
\left[
\begin{array}[l]{c}
s_k\\
v_k\\
a_k
\end{array}
\right]\enspace.
\end{align}
In the CA model, it is assumed that the acceleration $a$ is constant. The resulting system and input matrix is as follows
\begin{align}
\mat{A}^{(CA)}_k=\left[
\begin{array}[l]{ccc}
1 & T & \frac{1}{2} T^2\\
0 & 1 & T \\
0 & 0 & 1
\end{array}
\right],\quad \mat{B}^{(CA)}_k=\left[
\begin{array}[l]{c}
\frac{1}{2} T^2\\
T\\
1
\end{array}
\right]\enspace.
\end{align}
The covariance matrix of the process noise results analogously to the CV model, to
\begin{align}
\mat{Q}^{(CA)}_k &= 
          \left[
\begin{array}[l]{ccc}
\frac{1}{4} T^4 & \frac{1}{2} T^3 & \frac{1}{2} T^2\\
\frac{1}{2} T^3 & T^2 & T \\
\frac{1}{2} T^2 & T & 1
\end{array}
\right]
(\sigma^{(CA)}_w)^2 \enspace.
\end{align}

\bibliographystyle{IEEEtran}
\bibliography{Bibliography}
\end{document}